\documentstyle[prbbib,aps,preprint]{revtex}
\draft
\begin{document}

\title{Impact ionization 
in GaAs: a screened exchange density functional  approach}

\author{S. Picozzi\footnote{Corresponding author: {\tt
silvia.picozzi@aquila.infn.it}}}
\address{INFM -
Dip. Fisica,
Univ. L'Aquila, 67010 Coppito (L'Aquila), Italy }
\author{R.Asahi}
\address{Toyota Central R $\&$ D Labs., Inc., Nagakute, Japan}
\author{C.B.Geller}
\address{Bettis Atomic Power Laboratory, West Mifflin, PA 15122, USA}
\author{A.Continenza}
\address{INFM -
Dip. Fisica,
Univ. L'Aquila, 67010 Coppito (L'Aquila), Italy }
\author{A. J. Freeman}
\address{Department of Physics and Astronomy and Materials Research Center\\
Northwestern University, Evanston, IL 60208 (U.S.A.)\\}

\maketitle
\begin{abstract}
Results are presented of a fully ab-initio 
calculation of impact ionization rates in GaAs within the density
functional theory framework, using a screened--exchange formalism and
the highly precise all-electron
full-potential linearized augmented plane wave (FLAPW) method.
The calculated impact ionization
rates show a marked orientation dependence in { \bf k} space, indicating
the strong restrictions imposed by the conservation of energy and momentum.
This anisotropy diminishes  as the impacting 
electron energy increases. 
A Keldysh type fit performed on the energy--dependent rate shows a 
rather soft edge and a threshold energy  greater than the direct band gap.
The consistency with  available Monte Carlo and empirical
pseudopotential calculations shows the 
reliability of our approach and paves the way to ab-initio calculations of
pair production rates in new and more complex  materials.
\end{abstract}
Band-to-band impact ionization (I-I)
is a carrier-carrier scattering process in which an energetic carrier creates an
electron-hole pair through the excitation of a valence electron ($e^-$)
in the conduction
band\cite{landsberg,brennan1}. This process is fundamental in small high--speed
devices, 
both as a charge multiplication  (e.g., avalanche
photodiodes) and as a detrimental mechanism (e.g., field effect
transistors).   We present a fully first-principles approach
based on screened--exchange \cite{kleinman,seidl} density functional theory 
(DFT) \cite{dft} that allows 
a full  understanding of the basic mechanisms and physical quantities 
affecting the  I-I
process. Our final goal is to tune the rates 
according to technological requirements through 
 band structure ``engineering" 
 for simple and complex materials, 
without  need of  {\em ad--hoc} parameters or {\em pseudoatoms}.

All previous theoretical treatments have employed approximate band structures 
and matrix elements,
based on
${\bf k \cdot p}$, Monte Carlo (MC)\cite{hess,jung,brennan}
 or empirical pseudopotential (EPP)\cite{sy,schattke,hab_2}
 formalisms. 
To the best of our knowledge, the work herein reported 
is the {\em first fully ab--initio} calculation  of I-I rates.
We  show
  results for the most studied direct gap semiconductor,  GaAs.
A variety of I-I results obtained through different approaches
is available for GaAs, so that multiple comparisons can be made with results of
the current method.

We consider an $e^-$--initiated
I-I process (shown schematically in Fig. \ref{process}).
According to  Fermi's Golden rule, the 
 rate $r(n_1,{\bf k_1})$ at which the impacting $e^-$ in a $(n_1,{\bf k_1})$  state
 can produce I-I
is obtained as:

 \begin{eqnarray} r(n_1,{\bf k_1})=2 \frac{2\pi}{\hbar} \sum_{{n_2,n_3}}  
\int d^3 {\bf k_2}\int d^3 {\bf k_3 } |M|^2\\
\delta(E^{n_1}_{\bf k_1}+E^{n_2}_{\bf k_2}-E^{n_3}_{\bf k_3}-
E^{n_4}_{\bf k_1+k_2-k_3}),
\nonumber
\end{eqnarray}
where $n_1,n_2,n_3,n_4$  are  band indices ($n_i, i=1,....,7$ ) 
 and  ${\bf k_1,k_2,k_3}$  are    {\bf k}   points 
in  the {\em
full} Brillouin zone (BZ). 
 $\delta(E^{n_1}_{\bf k_1}+E^{n_2}_{\bf k_2}-E^{n_3}_{\bf k_3}-
E^{n_4}_{\bf k_1+k_2-k_3})$ shows the   energy conservation. 
The antisymmetrized screened Coulomb matrix element
is obtained by adding the probabilities in the singlet and triplet states
$|M|^2=\frac{1}{2}(|M_D|^2+|M_E|^2+|M_D-M_E|^2)$  where  $M_D$   
and   $M_E$    are {\em direct} and  {\em exchange} (obtained
from the direct term by exchanging final states) matrix elements.
Each direct matrix element is expressed as:
 \begin{eqnarray}
M_D = \frac{4\pi e^2}{\Omega}\sum_{\bf G_0,G_U} \delta({\bf
k_1+k_2-k_3-k_4+G_0})  \\
\frac{\rho_{n_3,{\bf k_3};n_1,{\bf k_1}}({\bf G_U})
 \rho_{n_4,{\bf k_4};n_2,{\bf k_2}}({\bf G_0-G_U})}
{\varepsilon(q)(|{\bf k_1 - k_3 + G_U}|^2+\lambda^2)}   
\nonumber
\end{eqnarray}

\noindent where the $\delta$   function shows the momentum ({\bf k})
conservation and    $\rho_{n_f,{\bf k_f};n_i,{\bf k_i}}({\bf G})$ 
   is the Fourier transform of  the  overlap
matrix of the wave functions, 
$\rho_{n_f,{\bf k_f};n_i,{\bf k_i}}({\bf r})=\Psi^*_{n_f,{\bf k_f}}({\bf r})
\Psi_{n_i,{\bf k_i}}({\bf r})$. The subscripts 
$i$ and $f$ denote initial and final states; 
 $e$  is the electron charge; 
$\Omega$  is the unit cell volume;  $q=|{\bf k_1 - k_3 + G_U}|$ is the 
momentum transfer and ${\bf G_0,G_U}$  are reciprocal lattice vectors. 

The
interaction between valence and conduction electrons is modeled
by a   Coulomb
potential screened through a $q$-dependent static model   dielectric
function $\varepsilon(q)$   
\cite{cappe},  particularly accurate for semiconductors.
The interaction between conduction electrons is modeled through a
Debye potential, commonly used for the screening of impurity
centers in semiconductors \cite{brennan1}. 
Both the
temperature $T$ and the  carrier  density in  the conduction band $n_0$ 
are taken into account through an inverse  Debye screening length
 given by $\lambda =\sqrt{\frac{4 \pi n_0
e^2}{K_B T}}$, where $K_B$  is the Boltzmann constant. 
Here, we used $T$ =
300 K and $n_0$ = 1$\cdot$10$^{16}$ cm$^{-3}$.
To carefully evaluate the matrix elements, 
we used  accurate wave
functions and 
band structures. In particular,
 to overcome the well--known shortcomings of the  
local density approximation (LDA) to  DFT
when dealing with excited states\cite{dft}, 
we performed self-consistent screened
exchange  (sX-LDA)\cite{kleinman,seidl,ryoji,apl} calculations, as
implemented within the highly precise 
full--potential linearized augmented
plane wave (FLAPW)\cite{FLAPW} method.
During the initial LDA (and the following 
sX-LDA) self--consistent iterations, we used a cut--off
equal to 9 Ry (6 Ry) in the wave vectors and $l\leq$8 ($l\leq$4) 
inside the
muffin--tin spheres chosen as $R_{MT}^{Ga}$ = $R_{MT}^{As}$ = 2.3 a.u.. The
summation over the irreducible
BZ was done using 4 special {\bf k}  points\cite{MP}.

 As shown by Seidl {\em et
al.}\cite{seidl}, the  sX-LDA approach can be recast within  a generalized
DFT  formalism 
in which the inclusion of a non--local sX functional
highly improves  the description of the
conduction band states compared with a bare LDA approach. This 
improvement is essential in the
present context, since the  transitions considered always involve conduction 
states. 
The many--fold integration over the full BZ was carried out using the
technique proposed by Sano and Yoshii\cite{sy}, 
based on a regular grid of  {\bf k}--points  
(with an interval length
  $\Delta k = (1/n) 2\pi/a$, $n$ = 10). 
As in previous inverstigations\cite{schattke,sy},   the energy
$\delta$ function  is approximated by a ``top-hat" function, {\em i.e.}
as a rectangle  $\Delta_E$ wide and $1/\Delta_E$ high 
(we used $\Delta_E$ = 0.2 eV\cite{schattke}).
Spin--orbit coupling and Umklapp  processes (${\bf G_0}\neq$ 0)
have been neglected; their inclusion in the formalism is the 
subject of ongoing work. 
We also have not considered: {\em i}) phonon--assisted transitions, that
 would relax
the {\bf k} conservation requirements among  the four involved electronic states
({\em i.e.}  {\bf k}  conservation would be satisfied through phonon 
participation) and {\em ii}) the influence of the 
electric field on the collision
term, {\em i.e.} the ``intra--collisional field effect"\cite{icfe}.

The results 
 for I-I rates 
initiated by electrons in the second lowest conduction band with wave vectors
along  [001]  and [111] are shown in Fig.\ref{anisotropy}. 
The insets show the band structure along the same symmetry lines
as in the main panels. Due to energy and  {\bf k} conservation constraints,
there is a  marked {\bf k}-space
 anisotropy in the I-I rate, 
  that is
 orders of magnitude higher
  along  [111] ($\Gamma$ to $L$)  
 than along [001] ($\Gamma$ to $X$). 
It is noted  that the second lowest conduction band in GaAs shows a 
decrease in energy from $\Gamma$ toward  $X$,
but
  increases with  increasing $|{\bf k}|$  in other directions, such as  
  [111]. 
  We therefore expect, and find (see Fig.\ref{anisotropy}(a)),  a 
``wave-vector anti-threshold" along the [001] axis
at which I-I 
no longer becomes possible.
Our calculated points are compared with those of other
calculations 
\cite{schattke} obtained with EPP that investigated the effect 
 on the predicted rate assuming two different band structures
  \cite{srivastava,cohen}. This comparison  makes clear
 the importance of employing an accurate band structure (such as  sX-LDA
 FLAPW), to 
 obtain
 reliable rates. The degree of agreement between these calculations and
  ours 
 is reasonable, given that  the band structure and some of the
 numerical approximations \cite{schattke}
 made in the evaluation of matrix elements  differ.

 The calculated I-I rates for GaAs are shown in Fig.\ref{impact_fin} (a)
 for processes
 initiated by electrons in the three lowest conduction bands vs. 
 their impacting 
energy.
The scattered points in the low--energy region reflect the strong
anisotropy already noted: carriers with the same impacting 
energy but different wave
vectors can have widely varying rates. However, this anisotropy diminishes
at higher energies, due to the greater ease with which
  {\bf k} and energy
conservation restrictions can be satisfied.
Moreover, it is of interest to
 show  an ``isotropic" I-I rate, that  depends only
on the impacting electron energy $E$ (solid line in Fig.  
\ref{impact_fin} (a)):

\begin{equation}
R(E) = \frac{\sum_{n_1}\int d^3{\bf k_1} \:
\delta[E^{n_1}_{\bf k_1}-E] r(n_1,{\bf k_1})}
{\sum_{n_1}\int d^3 {\bf k_1}\:
\delta[E^{n_1}_{\bf k_1}-E]}.
\end{equation}

\noindent The physical reason to obtain an isotropic
{\em energy} dependent rather than a
{\em wave-vector} dependent rate is that in most 
common technological
devices in which high electric fields  usually are present, 
carriers are  scattered by phonons,
so as to reach similar energies, but largely
different wavevectors.

Further,
our  $E$-dependent rate has been fitted using a ``Keldysh--type" formula
(dashed line in  Fig.  
\ref{impact_fin} (a)),
$ R(E) = P
[E-E_{th}]^a$  where  $E_{th}$  is the ``isotropic" threshold energy, {\em i.e.}
the minimum energy at which the carrier is able to excite an $e^-$ in the valence
band and, therefore, to initiate I-I. Here,
$P,a$ and  $E_{th}$ have been treated as fitting parameters.
An optimized linear regression procedure
 yields a fitted value $E_{th}^{fit}\sim$ 1.8 eV, whereas
our simulations indicate no ionization events for impacting energies lower
than $E_{th}^{calc}\sim$ 1.86 eV (considered  as
our ``real"
value for $E_{th}$). The excellent agreement between 
  $E_{th}^{fit}$ and $E_{th}^{calc}$ provides confidence in
the numerical fit. As a result,
the  threshold is slightly higher than the energy gap 
and in good agreement with the relation
$E_{th} = 3/2 E_{gap}$ obtained from a 
parabolic band structure with constant
 effective masses \cite{landsberg}. However,
we  believe  this agreement to be fortuitous, since we have 
shown previously
the importance of a careful treatment of band  anisotropy.
Moreover, it is noted that the value of $E_{th}$ is exact within
the uncertainty given by $\Delta_E$.
 From the fit, we obtained 
$P$ = 3.5x10$^{10} s^{-1}eV^{-a}$ and $a \sim$  5.8; the high  value
of $a$
reveals the 
 ``soft" character of the GaAs threshold. 
 
 In Table \ref{keldysh} and 
 Fig. \ref{impact_fin} (b) we compare the result of our fit 
 with other Keldysh--type fits available from  MC \cite{jung}
 or EPP \cite{schattke,hab_2} methods.
 Our {\em ab--initio} results are in 
 overall good agreement - especially at low energies - with two of the
 previous works \cite{hab_2,jung}. While the threshold energies are within a
 very similar range, there are some differences in the $P$ and $a$ values;
these can be ascribed mainly
  to details of band structures  
  and numerical methods employed  for evaluating the rates ({\em e.g.} many--fold
  integration schemes). 
 
 While it would be important to compare our results with  experiment,
 the I-I 
 rate is not directly comparable. Rather,
 the quantity commonly measured is the electron ionization coefficient 
 $\alpha(F)$ (in cm$^{-1}$) as a function of the applied
 field $F$. This coefficient is related to
   the inverse of the mean distance traveled by
  carrier prior to I-I; contrarily,
 our calculated rates (when summed with
 the phonon rates and normalized) give the probability (in s$^{-1}$)
  of an I-I event as a function of the carrier's energy. Therefore, the
  calculated I-I rates
 and the measured I-I coefficient  are not  easily related.
 For the  experiments, we refer to Bulman {\em et
 al.} \cite{bulman}, who measured $\alpha(F)$ in 
 (100) GaAs  in  a large number
 of different $p^+n$ structures using avalanche noise and photocurrent
 multiplication: their  electron ionization threshold is 1.7 eV.
 Therefore, the agreement with 
  our value of $\sim$ 1.8 eV
 supports the reliability of our
 procedure.
 
 Finally, we offer some information about the distribution of
final states, that is of great 
interest for transport simulations. In Fig. \ref{final} (a), 
 we plot
the average final  
electron energy vs  impacting electron energy along with
the best linear fit. Similar to
 the rates themselves, the
 scattering of points off the straight line
  is evident in the low energy region.  The average final electron energy
  is not necessarily equal, even for primary impacting electrons having the same
  initial energy, because of the strong restrictions imposed by  
  {\bf k} conservation. On the other hand,
 as the impacting energy increases the linear fit improves significantly.
Moreover, we plot in Fig.\ref{final} (b) the percentage of  transitions
that involve one or both final states in the symmetry point valleys 
($L,\Gamma,X$) vs the impacting electron
energies. As expected, 
at low energy, most (even 100 $\%$ at some energies)
of the transitions involve one or even both final states in the
$\Gamma$ valley. However,  as the impacting electron energy increases,
the other valleys (especially  the $L$ valley) become accessible
and the final states are generally more spread  over the BZ.
 

Work at Northwestern University was supported by the NSF (through the
N.U. Materials Research Center). We thank Dr. Wolfgang Mannstadt for early
discussions and Dr. Ilmun Ju for providing helpful
information.

\begin{table}
\caption{Fitting parameters for the rates shown in Fig. 3 (b), using
the Keldysh fit
formula  $R(E) = P
[E-E_{th}]^a$.} 
\begin{tabular}{|l|lll|}
& $E_{th}$ (eV)  & $a$ & $P$ (s$^{-1}$ eV$^{-a}$)\\ \hline \hline
This work & 1.8 & 5.8 & 3.5x10$^{10}$ \\ 
Ref.[\protect\onlinecite{hab_2}] & 1.89  & 5.2 & 1.4x10$^{11}$\\ 
Ref.[\protect\onlinecite{jung}] & 1.73 & 7.8 & 4.57x10$^{10}$ \\ 
Ref.[\protect\onlinecite{schattke}] & 2.1  & 4.0 & 2x10$^{12}$\\ 
\end{tabular}
\label{keldysh}
\end{table}

\begin{figure}
\caption{Electron-initiated  I-I process. The initial electrons
in the conduction and valence bands are in states 1 and 2, 
respectively; after
the transition, the final electrons 
in the conduction bands are in states 3 and 4.}
\label{process}
\end{figure}

\begin{figure}
\caption{Calculated ionization rates (in s$^{-1}$)
for impacting $e^-$ in the second
conduction band and wave vector along  ((a)
$\Gamma-X$
and (b) $\Gamma-L$) vs normalized wave vector (filled diamonds).
The open symbols show previous
results from Ref. \protect\onlinecite{schattke}.
In each panel, the inset shows the band
structure along the corresponding
symmetry lines (the zero of  energy  is set to the conduction band
minimum (CBM)).}
\label{anisotropy}
\end{figure}

\begin{figure}
\caption{(a) Ionization rates (in s$^{-1}$) for initial $e^-$ in
the three lowest
conduction bands vs their  energies (the zero of  energy  is set to the
CBM). Filled circles: wave vector dependent ionization rate
$R({\bf k})$;
solid line: energy dependent ionization rate $R(E)$; dashed line: 
Keldysh fit.
(b) Keldysh-type fits taken from this work (dashed bold line), Ref.
\protect\onlinecite{schattke} (thin solid line), Ref. 
\protect\onlinecite{hab_2} (dot--dashed line) and Ref.
 \protect\onlinecite{jung} (dotted line).}
\label{impact_fin}
\end{figure}

\begin{figure}
\caption{Panel (a): Average final $e^-$ energy (filled circles)
vs impacting $e^-$ energy (in eV)
 referred to the conduction band minimum (zero of  energy).
The solid line shows the best linear fit. Panel (b): Percentage of transitions
that involve final states at the symmetry point ($\Gamma$  (triangles),
$L$ (diamonds) and $X$ (circles)) valleys 
vs impacting $e^-$
energies (in eV).}
\label{final}
\end{figure}

\end{document}